\documentclass[ aps,%
12pt,%
floatfix,%
final,%
notitlepage,%
oneside,%
onecolumn,%
nobibnotes,%
nofootinbib,%
superscriptaddress,%
showpacs,%
]%
{revtex4}

\usepackage{graphicx}
\usepackage{amssymb}

\newcommand{\GeV}{\mbox{GeV}} \newcommand{\MeV}{\mbox{MeV}}
\newcommand{\Br}{\mathrm{Br}}
 \newcommand{\ps}{\mbox{ps}}

\newcommand{\M}{\mathcal{M}}
\newcommand{\R}{\mathcal{R}}

\begin{document}
 
\title{Radiative $B_c$ meson decays $B_c \to \gamma u\bar{d}$}

\author{A.K. Likhoded}
\email{Anatolii.Likhoded@ihep.ru}
\affiliation{Institute for High Energy Physics, Protvino, Russia}
\affiliation{Moscow Institute of Physics and Technology, Dolgoprundy, Russia}

\author{A.V. Luchinsky}
\email{Alexey.Luchinsky@ihep.ru}
\affiliation{Institute for High Energy Physics, Protvino, Russia}
\affiliation{SSC RF ITEP of NRC “Kurchatov Institute”}

\author{S.V. Poslavsky}
\email{stvlpos@mail.ru`}
\affiliation{Institute for High Energy Physics, Protvino, Russia}
\affiliation{SSC RF ITEP of NRC “Kurchatov Institute”}

\begin{abstract}
Inclusive radiative decays $B_c\to\gamma u\bar{d}$ are considered. It is shown that photon emission allows one to bypass the chiral suppression and leads the increase of the branching fraction by about four orders of magnitude. The estimates of exclusive decay branching fractions of $B_c$ meson into different sets of $\pi$ mesons are also given.
\end{abstract}


\maketitle

\section{Introduction}

$B_c$ meson $(\bar{b}c)$ takes a special place in the family of heavy quarkonia. In contrast to $(c\bar{c})$ and $(b\bar{b})$ systems the decays of ground $B_c$ state are necessary caused by weak interaction and can be divided onto three classes:
\begin{itemize}
 \item $\bar{b}$ quark decays with spectator $c$;
 \item $c$ quark decays with spectator $\bar{b}$;
 \item annihilation decays $B_c\to e^+\nu_e$, $c\bar{s}$, $u\bar{d}$, etc.
\end{itemize}
Some decays of the first type, e.g. $B_c\to \psi^{(')} +X$ with $X=\pi$, $3\pi$, $5\pi$, $e\nu$ (see \cite{Aaij:2014bla,LHCb:2012ag,Aaij:2013oya,Abulencia:2006zu,Aaltonen:2007gv}) and only one decay of the second type ($B_c\to B_s\pi$) are experimentally measured for today.

Annihilation decays are not yet observed. Total branching fraction of these channels is about $10\%$ \cite{Gershtein:1994jw} and the main contribution is given by  $B_c\to\tau\nu$ and $B_c\to c\bar{s}$ decays. Channels with light quarks in the final state are almost negligible due to chirality suppression. This suppression, however, can be bypassed if one consider radiative decays, e.g. $B_c\to u\bar{d}+\gamma$. The probabilities of these decays are suppressed by fine structure constant $\alpha$, but the chirality suppression factor $(m_{u,d}/M_{B_c})^2$ is absent. Our article is devoted to studies of such decays.

\section{Light Meson Production in Radiative $B_c$ decays}

Decays of $B_c$ meson with only light particles in the final state necessary require the weak annihilation of quark-antiquark pair $c\bar{b}$. The effective Lagrangian of this interaction has the form
\begin{eqnarray}
 \mathcal L_\mathrm{eff} &=& -\frac{G_F}{\sqrt{2}} V_{bc} V_{ud} 
  \left( \bar{b} \gamma_\mu (1-\gamma_5) c \right)
  \left( \bar{u} \gamma^\nu (1-\gamma_5) d \right).
\end{eqnarray}
Higher order QCD corrections can raise this expression by a factor of $a_1(m_c) \approx 1.14$ \cite{Buchalla:1995vs}.

\begin{figure}
 \includegraphics[width=0.8\textwidth]{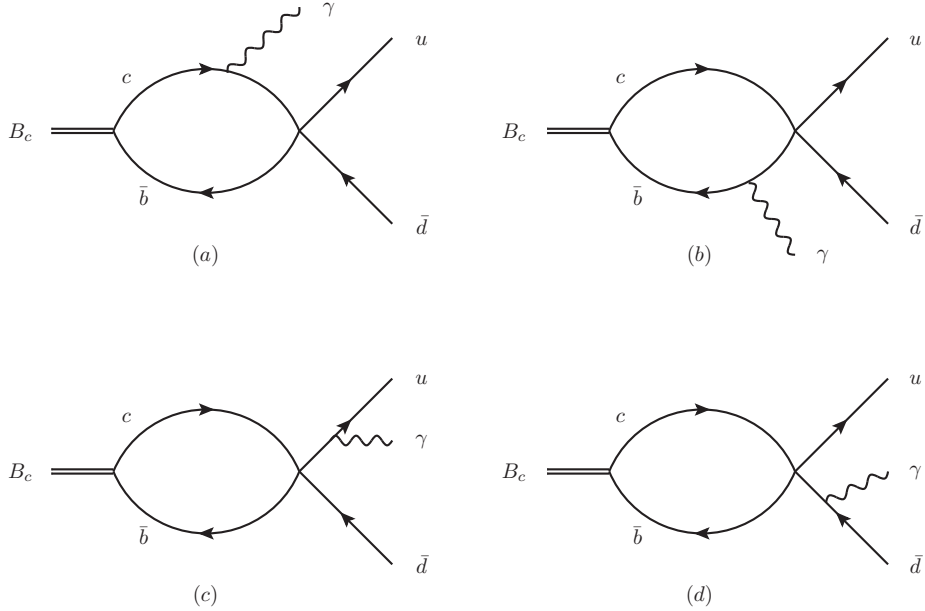}
 \caption{Feynman diagrams for $B_c \to \gamma u\bar{d}$ decay}
 \label{diags}
\end{figure}

Fig.~\ref{diags} shows Feynman diagrams corresponding to the inclusive radiative decay
\begin{eqnarray*}
B_c(P) &\to& \gamma(k) u(k_1)\bar{d}(k_2)
\end{eqnarray*}
in the considered approximation (particles momenta are shown in the parentheses).
It should be noted that at the same order the diagram with photon emission from virtual $W$ boson should also be included, but the corresponding amplitude is suppressed by additional Fermi constant, so in the following we will not take it into account.

The corresponding amplitude can be written in the form
\begin{eqnarray}
\M &=& \frac{4\pi\alpha G_F a_1}{\sqrt{2}} V_{bc} V_{ud}^* \epsilon^\mu \Big[
    H_{\mu\nu}(Q^2) L^\nu + H_\mu L^{\mu\nu}(Q^2)
  \Big],
  \label{eq:matr}
\end{eqnarray}
where $\epsilon_\mu$ is the final photon polarization vector, $Q^2=(k_1+k_2)^2$ is the squared invariant mass of quark-antiquark pair, and $H_{\mu\nu}$, $H_\nu$, $L_{\mu\nu}$, $L_\nu$ are the amplitudes of $B_c\to\gamma W$, $B_c\to W$, $W\to u\bar{d}\gamma$, $W\to u\bar{d}$ transitions respectively. The last two amplitudes are equal to
\begin{eqnarray}
 L_\nu &=& \delta^{ij} \bar{u}(k_1) \gamma_\nu d(k_2), \label{eq:Lm}\\
 L_{\mu\nu} &=& \delta^{ij} \left\{
  e_u \frac{1}{(k_1+k)^2} \bar{u}(k_1)\gamma_\mu (\hat{k}_1+\hat{k})\gamma_\nu (1+\gamma_5)d(k_2) +
\right.\nonumber\\  && \left.
  e_d \frac{1}{(k_2+k)^2} \bar{u}(k_1) \gamma_\nu(1+\gamma_5) (\hat{k}_2+\hat{k})\gamma_\mu d(k_2)
 \right\},\label{eq:Lmn}
\end{eqnarray}
where $i$, $j$ are quark's color indices, $e_u=2/3$, $e_d=-1/3$ are electric charges of $u$ and $d$ quarks, and we neglected the masses of final particles. The vertices of $B_c$ meson decays strongly depend on its internal structure and cannot be calculated in the framework of perturbation theory. It is convenient to write them in the general form
\begin{eqnarray}
 H_\nu &=& f_{B_c} P_\nu\label{eq:Hm}\\
 H_{\mu\nu}(Q^2) &=& A_g g_{\mu\nu} + A_{pp}(Q^2) P_\mu P_\nu + A_{kk}(Q^2) k_\mu k_\nu +
     A_{pk}(Q^2) P_\mu k_\nu + 
\nonumber \\ &&
     A_{kp}(Q^2) k_\mu P_\nu +
      i A_e(Q^2) e_{\mu\nu\alpha\beta} P^\alpha k^\beta,\label{eq:Hmn}
\end{eqnarray}
where $f_{B_c}$ is the leptonic constant of $B_c$ meson and $A_{g,pp,kk,pk,kp,e}$ are $B_c\to\gamma W$ decay form factors. Let us first consider the consequences of gauge invariance. In order to satisfy this condition  the vertices (\ref{eq:Lm}) -- (\ref{eq:Hmn}) should be connected by the following relations
\begin{eqnarray}
 k^\mu L_{\mu\nu} &=& (e_d-e_u) L_\nu,\qquad k^\mu H_{\mu\nu} = (e_c-e_b) H_\nu.
\end{eqnarray}
The validity of the first equality can easily be checked by direct calculations, and in order to satisfy the second one the following identities should hold:
\begin{eqnarray}
 (Pk) A_{pp}(Q^2) = (e_b - e_c) f_{B_c},\qquad A_g(Q^2) = -(Pk) A_{pk}(Q^2).
 \label{eq:ward}
\end{eqnarray}
It should be stresses that these identities are analogs of Ward identities in QED and do not depend on the model used for description of $B_c$ meson.

In order to obtain numerical results one should use some explicit expressions for presented in the vertex (\ref{eq:Hmn}) form factors. The simplest model is the assumption that heavy quarks stay at rest in $B_c$ meson rest frame, so their momenta are equal to
\begin{eqnarray}
 p_{b,c} &=& \frac{m_{b,c}}{M} P,\label{eq:pdelta}
\end{eqnarray}
where $m_{b,c}$ are constituent masses of $b$ and $c$ quarks, and total mass of $B_c$ meson is $M=m_b+m_c$. In this approximation the hadronization of $b\bar{c}$ pair is described by the projection operator
\begin{eqnarray}
 \Pi &=& \frac{\delta^{ij}}{2\sqrt{3}} \frac{1}{4m_bm_c}\frac{\psi(0)}{\sqrt{M}}
   (\hat{p_c} + m_c) (\hat{P}+M) \gamma_5 (\hat{p}_b - m_b),
\end{eqnarray}
where $i$, $j$ are quarks' color indices, and $\Psi(0)$ is $B_c$ meson wave function at the origin. Leptonic decay constant $f_{B_c}$ in this approximation is equal to \cite{Kiselev:2001xa,Kiselev:2003uk}
\begin{eqnarray*}
f_{B_{c}} & = & 2\sqrt{\frac{3}{M}}\Psi(0),
\end{eqnarray*}
where $\Psi(0)$ is a radial part of meson wave function at the origin. Then, the form factors are determined by the following expressions:
\begin{eqnarray*}
A_{g}^{\delta}\left(Q^{2}\right) & = & -\frac{\sqrt{3M}}{m_{b}m_{c}} a_1 \left|\Psi(0)\right|\left(e_{c}m_{b}-e_{b}m_{c}\right),\\
A_{pp}^{\delta}\left(Q^{2}\right) & = & 
      \frac{4\sqrt{3}}{\sqrt{M}} a_1\frac{\left(e_{b}-e_{c}\right)\left|\Psi(0)\right|}{M^{2}-Q^{2}},\\
A_{kk}^{\delta}\left(Q^{2}\right) & = & 0,\\
A_{pk}^{\delta}\left(Q^{2}\right) & = & A_{kp}^{\delta}\left(Q^{2}\right)=
    \frac{2\sqrt{3M}}{m_{b}m_{c}} a_1 \frac{\left(e_{c}m_{b}-e_{b}m_{c}\right)\left|\Psi(0)\right|}{M^{2}-Q^{2}},\\
A_{e}^{\delta}\left(Q^{2}\right) & = & \frac{2\sqrt{3M}}{m_{b}m_{c}} a_1 \frac{\left(e_{c}m_{b}+e_{b}m_{c}\right)\left|\Psi(0)\right|}{M^{2}-Q^{2}}.
\end{eqnarray*}
Differential and total widths of the considered decay in this approximation can be calculated analytically:
\begin{eqnarray}
 \frac{d^2\Gamma}{ds_k dQ^2} 	 &=& 
	\frac{3\alpha}{16\pi^2} 
	\left[	  a_1 V_{bc} V_{ud} \frac{f_{B_c} G_F}{m_c}\right]^2
	\frac{(e_b m_c + e_c m_b)^2}{M^3}
	\frac{Q^2}{(M^2-Q^2)^2} \times
	\nonumber\\ &&
	\left[
	  Q^4 + 2 s_2 Q^2 + s_2^2\left(1-\frac{m_c^2}{m_b^2}\right)-2M^2(s_2+Q^2)+M^4
	\right],\\
 \frac{d\Gamma}{dQ^2} 		 &=& 
	\frac{\alpha}{16\pi^2}
	\left[	  a_1 V_{bc} V_{ud} \frac{f_{B_c} G_F}{m_c m_b }\right]^2 
	\frac{(e_b m_c + m_c e_b)^2}{M^3}
	(m_b^2 + m_c^2)
	Q^2 (M^2-Q^2)
	,\\
 \Gamma  			 &=& 
	\frac{\alpha}{96\pi^2}\left[\frac{a_1 V_{bc}V_{ud}f_{B_c}G_F}{m_b m_c}\right]^2
	M^3 (e_b m_c+e_c m_b)^2,
\end{eqnarray}
where $s_2=(P-k_2)^2$. It is interesting to note that these expressions are free from collinear and infrared singularities and tend to zero in $Q^2 \to 0$ and $Q^2 \to M^2$ limits. This behavior is explained by mentioned above chirality suppression.

\begin{figure}
\includegraphics[width=\textwidth]{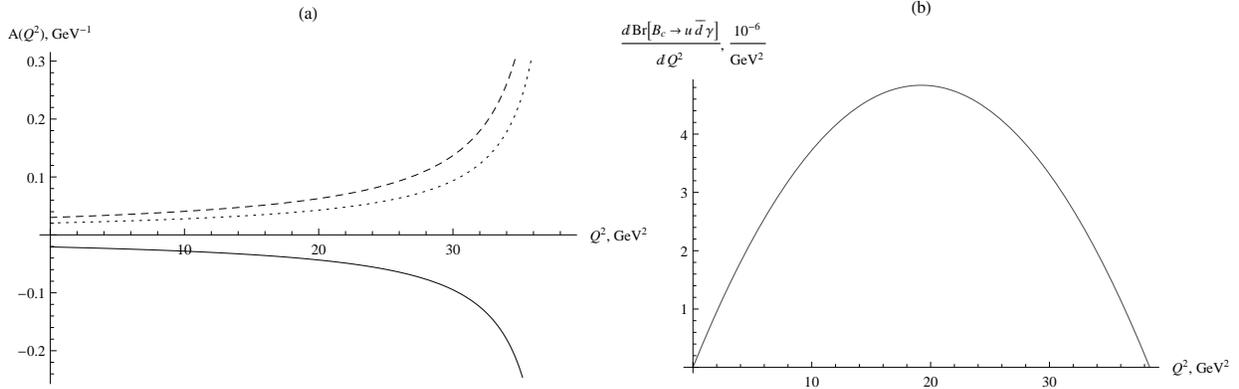}
 \caption{Left pabel: $B_c\to\gamma W$ vertex form factors. Solid, dashed and dotted lines correspond to $A_{pp}$, $A_{pk}$, and $A_e$ form factors respectively. The others are connected with them by the Ward identities (\ref{eq:ward}); Right panel: transferred momentum distribution of $B_c\to\gamma u\bar{d}$ decay branching fraction
\label{fig:ff}}
\end{figure}

In our paper we use the following values of model parameters:
\begin{eqnarray*}
 &&M_{B_c} = 6.2\,\GeV,\qquad m_b=4.5\,\GeV, \qquad m_c=1.7\,\GeV\\
  &&f_{B_c} = 400\,\MeV, \qquad V_{bc}=0.045,\qquad \tau_{B_c} = 0.452\,\ps
\end{eqnarray*}
Transferred momentum distributions of $B_c\to\gamma W$ form factors and $B_c\to\gamma u\bar{d}$ branching fraction are  shown in left and right panels of Fig.\ref{fig:ff}. Integrated branching fraction of the inclusive decay $B_c\to\gamma u\bar{d}$ with this choice of the parameters is equal to
\begin{eqnarray*}
 \Br[B_c\to \gamma u \bar{d}] &=& 1.3\times 10^{-4}.
\end{eqnarray*}

Let us now consider the case of exlusive decays. According to the factorization theorem, for the exclusive decay $B_c\to\gamma\R$ the distribution of the branching fraction over the squared invariant mass of light mesons system $Q^2$  is connected with the width of the inclusive decay $B_c \to \gamma u\bar{d}$ \cite{Schael:2005am,Likhoded:2009ib}:
\begin{eqnarray}
 \frac{d\Gamma(B_c\to\gamma\R)}{dQ^2} &=& \int\frac{dQ^2}{2\pi} \frac{d\Gamma(B_c\to\gamma u\bar{d})}{dQ^2} \rho^\R(Q^2),
\end{eqnarray}
where the spectral function $\rho^\R$ describes the hadronization of $u\bar{d}$ pair into final state $\R$.

\begin{figure}
 \includegraphics[width=0.9\textwidth]{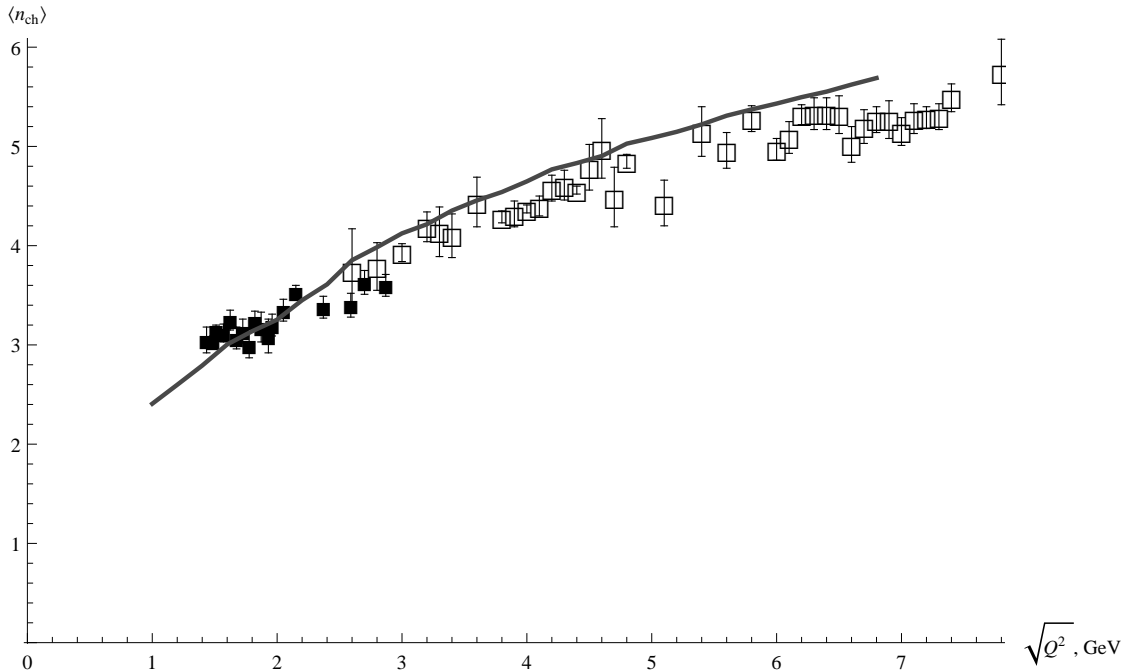}
 \caption{Charged particle multiplicity calculated using Pythia generator in comparison with experimental data \cite{Bacci:1979ab,Siegrist:1981zp}}
\label{fig:nch}
\end{figure}

\begin{figure}
 \includegraphics[width=\textwidth]{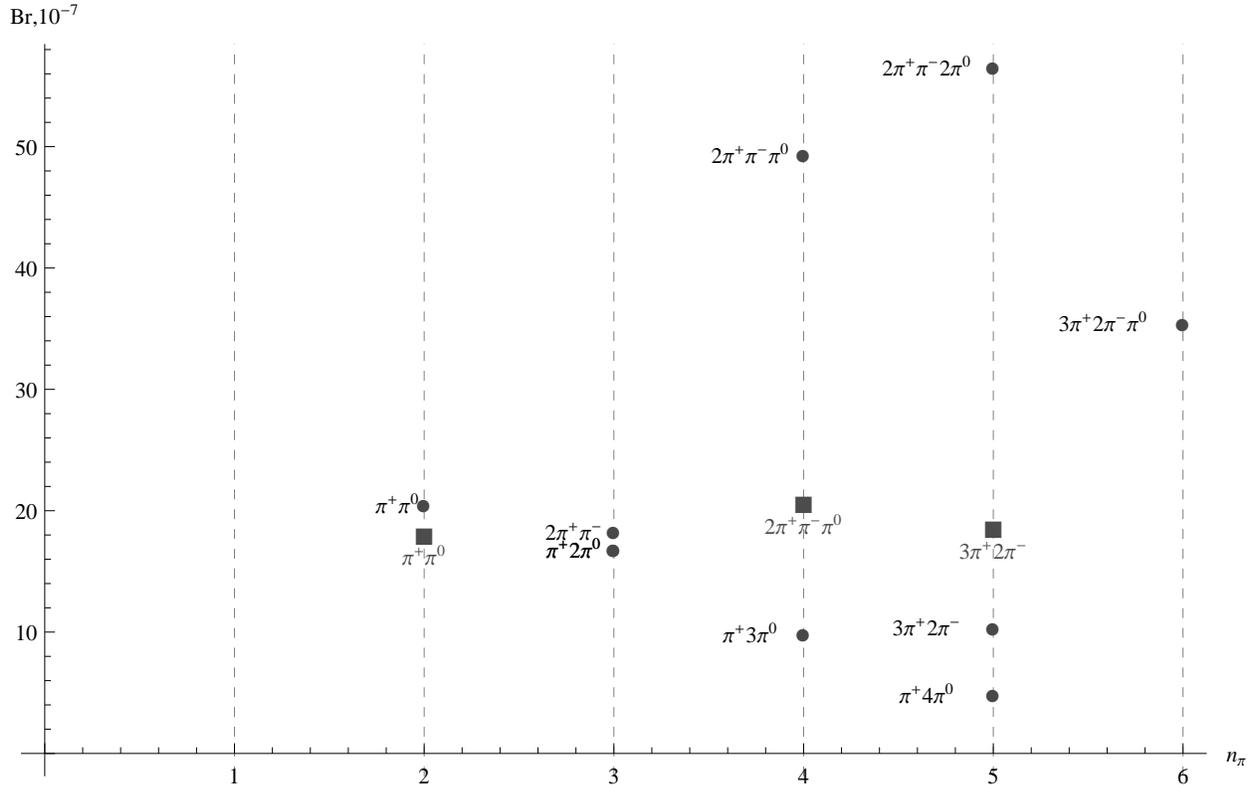}
 \caption{
 Branching fractions of exclusive $B_c\to\gamma\R$ decays for different models of $u\bar{d}$ pair hadronization. Symbols ``$\bullet$'' 
and ``$\blacksquare$''
corrspond to predictions based on Pythia generator and $\tau$ lepton decays analysis respectively
}
\label{fig:plotSP}
\end{figure}

These spectral functions can be determined, for example, from the analysis of the exclusive decays $\tau\to\nu_\tau\R$, as it was done in \cite{Likhoded:2009ib}. The drawback of such approach is that the kinematical region of $Q^2$ in our case is larger than in the case of $\tau$ lepton decays. An alternative approach is to consider hadronization of quark-antiquark pair by mesons in Pythia event generator \cite{Sjostrand:2007gs}. Using this method one can determine the spectral functions over the whole kinematic region. As it seen from  Fig.~\ref{fig:nch} such approach describes well the experimental data for mean charged particles multiplicity. It should be noted, however, that this approach does not allow one to control the conservation of some quantum numbers, e.g. isospin or charge parity of the final state. For this reason presented below results should be considered as rough estimated. The values of integrated branching fractions are presented in Fig.\ref{fig:plotSP}.

\section{Conclusion\label{sec:Conclusion}}

In the presented paper the production of light mesons in $B_c$ decays with emission of additional photon is considered.

It is well known that at ledaing order without photon emission such decays are strongly suppressed by chirality conservation. According to presented in \cite{Gershtein:1994jw} results the branching fraction of $B_c\to u\bar{d}$ decay is about $10^{-8}$ (such suppression is absent if charged Higgs boson contribution \cite{Lebedev:2000ix,Park:2006gk,Raidal:2008jk} are taken into account, so considered decay can be used to set an upper bound on its mass). In our paper we show, that consideration of higher order processes $B_c\to u\bar{d}\gamma$ allows one to bypass this rule, so the branching fractions increase by about four orders of magnitude. It is clear also, that the same effect can be observed in recently detected $B_s\to\mu^+\mu^-$ decay \cite{Aditya:2012im,Aaij:2013aka,Chatrchyan:2013bka}.

In our work we calculate analytical expressions for differential and integrated branching fractions of inclusive $B_c\to u\bar{d}\gamma$ decay and present numerical results. In addition, using experimental data on $\tau$-lepton decays and charged meson multiplicities we obtain the estimates for branching fractions of exclusive decays.

It should be noted that in our calculations the leading order nonrelativistic Quantum Chromodynamic was used, so we neglected the intrinsic motion of quarks in $B_c$ meson. One can expect, however, that, similar to double charmonium production in electron-positron annihilation \cite{Braguta:2005kr,Abe:2002rb} and bottomonia decays \cite{Braguta:2009df,Braguta:2009xu,Shen:2012ei}, this effect could lead to significant increase of the branching fractions. 

The authors would like to thank V.V. Kiselev for fruitful discussions. The work was financially supported by RFBR (\#14-02-00096 A) and  FRRC grants.


\begin{thebibliography}{25}
\expandafter\ifx\csname natexlab\endcsname\relax\def\natexlab#1{#1}\fi
\expandafter\ifx\csname bibnamefont\endcsname\relax
  \def\bibnamefont#1{#1}\fi
\expandafter\ifx\csname bibfnamefont\endcsname\relax
  \def\bibfnamefont#1{#1}\fi
\expandafter\ifx\csname citenamefont\endcsname\relax
  \def\citenamefont#1{#1}\fi
\expandafter\ifx\csname url\endcsname\relax
  \def\url#1{\texttt{#1}}\fi
\expandafter\ifx\csname urlprefix\endcsname\relax\def\urlprefix{URL }\fi
\providecommand{\bibinfo}[2]{#2}
\providecommand{\eprint}[2][]{\url{#2}}

\bibitem[{\citenamefont{Aaij et~al.}(2014)}]{Aaij:2014bla}
\bibinfo{author}{\bibfnamefont{R.}~\bibnamefont{Aaij}} \bibnamefont{et~al.}
  (\bibinfo{collaboration}{LHCb collaboration}),
  \eprint{arXiv:1404.0287}.

\bibitem[{\citenamefont{Aaij et~al.}(2012)}]{LHCb:2012ag}
\bibinfo{author}{\bibfnamefont{R.}~\bibnamefont{Aaij}} \bibnamefont{et~al.}
  (\bibinfo{collaboration}{LHCb Collaboration}),
  \bibinfo{journal}{Phys.Rev.Lett.} \textbf{\bibinfo{volume}{108}},
  \bibinfo{pages}{251802} (\bibinfo{year}{2012}), \eprint{arXiv:1204.0079}.

\bibitem[{\citenamefont{Aaij et~al.}(2013{\natexlab{a}})}]{Aaij:2013oya}
\bibinfo{author}{\bibfnamefont{R.}~\bibnamefont{Aaij}} \bibnamefont{et~al.}
  (\bibinfo{collaboration}{LHCb collaboration}), \bibinfo{journal}{Phys.Rev.}
  \textbf{\bibinfo{volume}{D87}}, \bibinfo{pages}{071103}
  (\bibinfo{year}{2013}{\natexlab{a}}), \eprint{arXiv:1303.1737}.

\bibitem[{\citenamefont{Abulencia et~al.}(2006)}]{Abulencia:2006zu}
\bibinfo{author}{\bibfnamefont{A.}~\bibnamefont{Abulencia}}
  \bibnamefont{et~al.} (\bibinfo{collaboration}{CDF Collaboration}),
  \bibinfo{journal}{Phys.Rev.Lett.} \textbf{\bibinfo{volume}{97}},
  \bibinfo{pages}{012002} (\bibinfo{year}{2006}), \eprint{hep-ex/0603027}.

\bibitem[{\citenamefont{Aaltonen et~al.}(2008)}]{Aaltonen:2007gv}
\bibinfo{author}{\bibfnamefont{T.}~\bibnamefont{Aaltonen}} \bibnamefont{et~al.}
  (\bibinfo{collaboration}{CDF Collaboration}),
  \bibinfo{journal}{Phys.Rev.Lett.} \textbf{\bibinfo{volume}{100}},
  \bibinfo{pages}{182002} (\bibinfo{year}{2008}), \eprint{arXiv:0712.1506}.

\bibitem[{\citenamefont{Gershtein et~al.}(1995)\citenamefont{Gershtein,
  Kiselev, Likhoded, and Tkabladze}}]{Gershtein:1994jw}
\bibinfo{author}{\bibfnamefont{S.}~\bibnamefont{Gershtein}},
  \bibinfo{author}{\bibfnamefont{V.}~\bibnamefont{Kiselev}},
  \bibinfo{author}{\bibfnamefont{A.}~\bibnamefont{Likhoded}}, \bibnamefont{and}
  \bibinfo{author}{\bibfnamefont{A.~a.} \bibnamefont{Tkabladze}},
  \bibinfo{journal}{Phys.Usp.} \textbf{\bibinfo{volume}{38}},
  \bibinfo{pages}{1} (\bibinfo{year}{1995}), \eprint{hep-ph/9504319}.

\bibitem[{\citenamefont{Buchalla et~al.}(1996)\citenamefont{Buchalla, Buras,
  and Lautenbacher}}]{Buchalla:1995vs}
\bibinfo{author}{\bibfnamefont{G.}~\bibnamefont{Buchalla}},
  \bibinfo{author}{\bibfnamefont{A.~J.} \bibnamefont{Buras}}, \bibnamefont{and}
  \bibinfo{author}{\bibfnamefont{M.~E.} \bibnamefont{Lautenbacher}},
  \bibinfo{journal}{Rev.Mod.Phys.} \textbf{\bibinfo{volume}{68}},
  \bibinfo{pages}{1125} (\bibinfo{year}{1996}), \eprint{hep-ph/9512380}.

\bibitem[{\citenamefont{Kiselev et~al.}(2002)\citenamefont{Kiselev, Likhoded,
  Pakhomova, and Saleev}}]{Kiselev:2001xa}
\bibinfo{author}{\bibfnamefont{V.}~\bibnamefont{Kiselev}},
  \bibinfo{author}{\bibfnamefont{A.}~\bibnamefont{Likhoded}},
  \bibinfo{author}{\bibfnamefont{O.}~\bibnamefont{Pakhomova}},
  \bibnamefont{and} \bibinfo{author}{\bibfnamefont{V.}~\bibnamefont{Saleev}},
  \bibinfo{journal}{Phys.Rev.} \textbf{\bibinfo{volume}{D65}},
  \bibinfo{pages}{034013} (\bibinfo{year}{2002}), \eprint{hep-ph/0105190}.

\bibitem[{\citenamefont{Kiselev}(2004)}]{Kiselev:2003uk}
\bibinfo{author}{\bibfnamefont{V.}~\bibnamefont{Kiselev}},
  \bibinfo{journal}{Central Eur.J.Phys.} \textbf{\bibinfo{volume}{2}},
  \bibinfo{pages}{523} (\bibinfo{year}{2004}), \eprint{hep-ph/0304017}.

\bibitem[{\citenamefont{Schael et~al.}(2005)}]{Schael:2005am}
\bibinfo{author}{\bibfnamefont{S.}~\bibnamefont{Schael}} \bibnamefont{et~al.}
  (\bibinfo{collaboration}{ALEPH Collaboration}), \bibinfo{journal}{Phys.Rept.}
  \textbf{\bibinfo{volume}{421}}, \bibinfo{pages}{191} (\bibinfo{year}{2005}),
  \eprint{hep-ex/0506072}.

\bibitem[{\citenamefont{Likhoded and Luchinsky}(2010)}]{Likhoded:2009ib}
\bibinfo{author}{\bibfnamefont{A.}~\bibnamefont{Likhoded}} \bibnamefont{and}
  \bibinfo{author}{\bibfnamefont{A.}~\bibnamefont{Luchinsky}},
  \bibinfo{journal}{Phys.Rev.} \textbf{\bibinfo{volume}{D81}},
  \bibinfo{pages}{014015} (\bibinfo{year}{2010}), \eprint{arXiv:0910.3089}.

\bibitem[{\citenamefont{Bacci et~al.}(1979)\citenamefont{Bacci, Baldini~Celio,
  Battistoni, Bollini, Capon et~al.}}]{Bacci:1979ab}
\bibinfo{author}{\bibfnamefont{C.}~\bibnamefont{Bacci}},
  \bibinfo{author}{\bibfnamefont{R.}~\bibnamefont{Baldini~Celio}},
  \bibinfo{author}{\bibfnamefont{G.}~\bibnamefont{Battistoni}},
  \bibinfo{author}{\bibfnamefont{D.}~\bibnamefont{Bollini}},
  \bibinfo{author}{\bibfnamefont{G.}~\bibnamefont{Capon}},
  \bibnamefont{et~al.}, \bibinfo{journal}{Phys.Lett.}
  \textbf{\bibinfo{volume}{B86}}, \bibinfo{pages}{234} (\bibinfo{year}{1979}).

\bibitem[{\citenamefont{Siegrist et~al.}(1982)\citenamefont{Siegrist,
  Schwitters, Alam, Boyarski, Breidenbach et~al.}}]{Siegrist:1981zp}
\bibinfo{author}{\bibfnamefont{J.}~\bibnamefont{Siegrist}},
  \bibinfo{author}{\bibfnamefont{R.}~\bibnamefont{Schwitters}},
  \bibinfo{author}{\bibfnamefont{M.}~\bibnamefont{Alam}},
  \bibinfo{author}{\bibfnamefont{A.}~\bibnamefont{Boyarski}},
  \bibinfo{author}{\bibfnamefont{M.}~\bibnamefont{Breidenbach}},
  \bibnamefont{et~al.}, \bibinfo{journal}{Phys.Rev.}
  \textbf{\bibinfo{volume}{D26}}, \bibinfo{pages}{969} (\bibinfo{year}{1982}).

\bibitem[{\citenamefont{Sjostrand et~al.}(2008)\citenamefont{Sjostrand, Mrenna,
  and Skands}}]{Sjostrand:2007gs}
\bibinfo{author}{\bibfnamefont{T.}~\bibnamefont{Sjostrand}},
  \bibinfo{author}{\bibfnamefont{S.}~\bibnamefont{Mrenna}}, \bibnamefont{and}
  \bibinfo{author}{\bibfnamefont{P.~Z.} \bibnamefont{Skands}},
  \bibinfo{journal}{Comput.Phys.Commun.} \textbf{\bibinfo{volume}{178}},
  \bibinfo{pages}{852} (\bibinfo{year}{2008}), \eprint{arXiv:0710.3820}.

\bibitem[{\citenamefont{Lebedev et~al.}(2000)\citenamefont{Lebedev, Loinaz, and
  Takeuchi}}]{Lebedev:2000ix}
\bibinfo{author}{\bibfnamefont{O.}~\bibnamefont{Lebedev}},
  \bibinfo{author}{\bibfnamefont{W.}~\bibnamefont{Loinaz}}, \bibnamefont{and}
  \bibinfo{author}{\bibfnamefont{T.}~\bibnamefont{Takeuchi}},
  \bibinfo{journal}{Phys.Rev.} \textbf{\bibinfo{volume}{D62}},
  \bibinfo{pages}{055014} (\bibinfo{year}{2000}), \eprint{hep-ph/0002106}.

\bibitem[{\citenamefont{Park}(2006)}]{Park:2006gk}
\bibinfo{author}{\bibfnamefont{J.-h.} \bibnamefont{Park}},
  \bibinfo{journal}{JHEP} \textbf{\bibinfo{volume}{0610}}, \bibinfo{pages}{077}
  (\bibinfo{year}{2006}), \eprint{hep-ph/0607280}.

\bibitem[{\citenamefont{Raidal et~al.}(2008)\citenamefont{Raidal, van~der
  Schaaf, Bigi, Mangano, Semertzidis et~al.}}]{Raidal:2008jk}
\bibinfo{author}{\bibfnamefont{M.}~\bibnamefont{Raidal}},
  \bibinfo{author}{\bibfnamefont{A.}~\bibnamefont{van~der Schaaf}},
  \bibinfo{author}{\bibfnamefont{I.}~\bibnamefont{Bigi}},
  \bibinfo{author}{\bibfnamefont{M.}~\bibnamefont{Mangano}},
  \bibinfo{author}{\bibfnamefont{Y.~K.} \bibnamefont{Semertzidis}},
  \bibnamefont{et~al.}, \bibinfo{journal}{Eur.Phys.J.}
  \textbf{\bibinfo{volume}{C57}}, \bibinfo{pages}{13} (\bibinfo{year}{2008}),
  \eprint{arXiv:0801.1826}.

\bibitem[{\citenamefont{Aditya et~al.}(2013)\citenamefont{Aditya, Healey, and
  Petrov}}]{Aditya:2012im}
\bibinfo{author}{\bibfnamefont{Y.}~\bibnamefont{Aditya}},
  \bibinfo{author}{\bibfnamefont{K.}~\bibnamefont{Healey}}, \bibnamefont{and}
  \bibinfo{author}{\bibfnamefont{A.~A.} \bibnamefont{Petrov}},
  \bibinfo{journal}{Phys.Rev.} \textbf{\bibinfo{volume}{D87}},
  \bibinfo{pages}{074028} (\bibinfo{year}{2013}), \eprint{arXiv:1212.4166}.

\bibitem[{\citenamefont{Aaij et~al.}(2013{\natexlab{b}})}]{Aaij:2013aka}
\bibinfo{author}{\bibfnamefont{R.}~\bibnamefont{Aaij}} \bibnamefont{et~al.}
  (\bibinfo{collaboration}{LHCb collaboration}),
  \bibinfo{journal}{Phys.Rev.Lett.} \textbf{\bibinfo{volume}{111}},
  \bibinfo{pages}{101805} (\bibinfo{year}{2013}{\natexlab{b}}),
  \eprint{arXiv:1307.5024}.

\bibitem[{\citenamefont{Chatrchyan et~al.}(2013)}]{Chatrchyan:2013bka}
\bibinfo{author}{\bibfnamefont{S.}~\bibnamefont{Chatrchyan}}
  \bibnamefont{et~al.} (\bibinfo{collaboration}{CMS Collaboration}),
  \bibinfo{journal}{Phys.Rev.Lett.} \textbf{\bibinfo{volume}{111}},
  \bibinfo{pages}{101804} (\bibinfo{year}{2013}), \eprint{arXiv:1307.5025}.

\bibitem[{\citenamefont{Braguta et~al.}(2005)\citenamefont{Braguta, Likhoded,
  and Luchinsky}}]{Braguta:2005kr}
\bibinfo{author}{\bibfnamefont{V.}~\bibnamefont{Braguta}},
  \bibinfo{author}{\bibfnamefont{A.}~\bibnamefont{Likhoded}}, \bibnamefont{and}
  \bibinfo{author}{\bibfnamefont{A.}~\bibnamefont{Luchinsky}},
  \bibinfo{journal}{Phys.Rev.} \textbf{\bibinfo{volume}{D72}},
  \bibinfo{pages}{074019} (\bibinfo{year}{2005}), \eprint{hep-ph/0507275}.

\bibitem[{\citenamefont{Abe et~al.}(2002)}]{Abe:2002rb}
\bibinfo{author}{\bibfnamefont{K.}~\bibnamefont{Abe}} \bibnamefont{et~al.}
  (\bibinfo{collaboration}{Belle Collaboration}),
  \bibinfo{journal}{Phys.Rev.Lett.} \textbf{\bibinfo{volume}{89}},
  \bibinfo{pages}{142001} (\bibinfo{year}{2002}), \eprint{hep-ex/0205104}.

\bibitem[{\citenamefont{Braguta et~al.}(2009)\citenamefont{Braguta, Likhoded,
  and Luchinsky}}]{Braguta:2009df}
\bibinfo{author}{\bibfnamefont{V.}~\bibnamefont{Braguta}},
  \bibinfo{author}{\bibfnamefont{A.}~\bibnamefont{Likhoded}}, \bibnamefont{and}
  \bibinfo{author}{\bibfnamefont{A.}~\bibnamefont{Luchinsky}},
  \bibinfo{journal}{Phys.Rev.} \textbf{\bibinfo{volume}{D80}},
  \bibinfo{pages}{094008} (\bibinfo{year}{2009}), \eprint{arXiv:0902.0459}.

\bibitem[{\citenamefont{Braguta and Kartvelishvili}(2010)}]{Braguta:2009xu}
\bibinfo{author}{\bibfnamefont{V.}~\bibnamefont{Braguta}} \bibnamefont{and}
  \bibinfo{author}{\bibfnamefont{V.}~\bibnamefont{Kartvelishvili}},
  \bibinfo{journal}{Phys.Rev.} \textbf{\bibinfo{volume}{D81}},
  \bibinfo{pages}{014012} (\bibinfo{year}{2010}), \eprint{arXiv:0907.2772}.

\bibitem[{\citenamefont{Shen et~al.}(2012)\citenamefont{Shen, Yuan, and
  Iijima}}]{Shen:2012ei}
\bibinfo{author}{\bibfnamefont{C.}~\bibnamefont{Shen}},
  \bibinfo{author}{\bibfnamefont{C.}~\bibnamefont{Yuan}}, \bibnamefont{and}
  \bibinfo{author}{\bibfnamefont{T.}~\bibnamefont{Iijima}}
  (\bibinfo{collaboration}{Belle Collaboration}), \bibinfo{journal}{Phys.Rev.}
  \textbf{\bibinfo{volume}{D85}}, \bibinfo{pages}{071102}
  (\bibinfo{year}{2012}), \eprint{arXiv:1203.0368}.

\end{thebibliography}

\end{document}